# Can Fires, Night Lights, and Mobile Phones reveal behavioral fingerprints useful for Development?

David Pastor-Escuredo[1,2], Thierry Savy[3] and Miguel A. Luengo-Oroz[1,2]

[1] Biomedical Image Technologies, DIE- ETSIT, Universidad Politécnica de Madrid, CEI Moncloa UPM-UCM, Madrid, Spain.

[2] Biomedical Research Networking Center in Bioengineering, Biomaterials and Nanomedicine, CIBER-BBN, Spain

[3] Complex Systems Institute of Paris Ile-de-France, ISC-PIF, Paris, France

**ABSTRACT**

*Fires, lights at night and mobile phone activity have been separately used as proxy indicators of human activity with high potential for measuring human development. In this preliminary report, we develop some tools and methodologies to identify and visualize relations among remote sensing datasets containing fires and night lights information with mobile phone activity in Cote D'Ivoire from December 2011 to April 2012.*

## 1. Introduction

Fires, luminosity from lights at night and mobile phone activity are three ubiquous signals which can serve as proxy indicators of human activity. Recent literature shows that the ability to measure these signals can be used to understand patterns reflecting economic development and modulating factors as violent conflicts.

*Fire data*

We can identify and quantify fires at a global scale using remote sensing (eg, data from MODIS imaging system). Fire detection data is used for forest and agricultural monitoring, climate change and air quality modeling (Davies 2009). Fire detection might also be used as an input into early warning systems to flag potential human rights violations or humanitarian emergencies. For instance, it has been used to identify burning campaigns in human settlements in Darfur (Sudan) during periods of ethnic violence (Bromley 2010). In another study in Kenya, the United Nations used satellite fire imagery to locate areas where violence had potentially occurred (Anderson 2008).

*Luminosity data*

Another type of satellite data that can provide useful information of conflict zones is imagery of luminosity from lights at night visible from space (eg. data from DMSP-OLS sensors). Changes in the lights at night signature in cities from Russia and Georgia between 1992 and 2009 were measured to detect the effects of war in the Caucasus region, with the potential application of corroborating reports of unknown quality that emanate from war zones (Witmer 2011)]. Luminosity measurements have also been shown as a way to improve the quality of socioeconomic indicators in developing countries whose standard statistic data sources may be weak, for example countries with no recent population or economic censuses. In particular, luminosity has been correlated against the gross domestic product of several countries during the period 1992-2008 (Chen 2011).

*Mobile Phone Data*

Call detail records from mobile phone activity have recently been proposed as a potential and promising source of information to understand human behavior that might lead to proxy indicators for populations' well-being. For some information, data extracted from mobile phone activity may provide a less expensive alternative to field surveys and may be useful to helping policy makers monitor existing programs in remote locations. More precisely, recent breakthrough research has explored how calling patterns such as call reciprocity and call diversity can be used to detect the socioeconomic status of populations. This kind of research has been developed using data from UK (Eagle 2010), Rwanda (Blumenstock 2010) and Latin America (Frias-Martinez 2012). Mobile phone data has also been used to track population movements following natural disasters or economic shocks. After Haiti's earthquake, a study in areas of high mobile use showed that mobile data can rapidly provide estimates of population movements during disasters (Bengtsson 2011, Lu 2012). Another recent study in Kenya showed correlations between mobile phone data used to measure populations movements on weekly, monthly, and annual time scales with data on change in residence from the national census conducted during the same time period (Wesolowski 2013). These methods have also been used to understand migration in urban settlements (slums) in Kenya o infer information on informal employment (Wesolowski 2010). In another example, spatially explicit mobile phone data and malaria prevalence information from Kenya have been used to identify the dynamics of human carriers that drive parasite importation between regions (Wesolowski 2012).

In this report, we present our preliminary research devoted to understand how to merge, visualize and analyze information from mobile phone call detail records with remote sensing - fire and light - data in order to discover potential applications for development. First, assuming the hypothesis that mobile phone activity can be used to understand human behaviours, we will explore some relationships between mobile phone activity, lights at night, and locations with significant fires. Then, we present our on-going efforts to develop a visualization software able to integrate data contained in the call detail records with mapping and remote sensing information (in this case, fire and light). The initial purpose of these tools is to better understand what types of questions related to development is possible to ask with these types of data, and which are the right tools that can be used to answer them. Therefore, we close the report with some open questions and ideas that we hope can be used to explore new research ventures.

## 2. Data

### 2.1 D4D Data

The mobile phone datasets are based on anonymized Call Detail Records (CDR) of phone calls and SMS exchanges between a subsample of Orange's customers in Ivory Coast. The datasets were made available under the Data for Development Challenge and their description can be found [here](#) (Blondel 2012).The data covers a total of 3600 hours between December 1, 2011 and April 28, 2012. Due to technical reasons data is missing for total period of about 100 hours. In particular, in this research, we have used two types of data:

*a- Antenna-to-antenna traffic.* For 1231 antennas with a precise geographic location, the number of calls as well as the duration of calls between any pair of antennas has been aggregated hour by hour. This data is available for the entire observation period and represents a subset of the communications between Orange customers.

*b- Individual Trajectories: High Spatial Resolution Data.* Individual movement trajectories can be approximated from the geographic location of the cell phone antennas during calls. This dataset contains high resolution trajectories of 50000 randomly sampled individuals over two-week periods. To protect privacy new random identifiers (individuals) are chosen in every time period. Time stamps are rounded to the minute.

### 2.2 Fire data

Using the NASA FIRMS resource that provides fire detection at a global scale (earthdata.nasa.gov/firms), we have filtered all the fires occurring in a geographical bounding box that covers Cote d'Ivore. In total, we have found 59469 fires between December 1, 2011 and April 28, 2012 in the selected region (see Fig1.a). The fires are detected using data from the MODIS instrument, on board NASA's Aqua and Terra satellites (Davies 2009). The satellites take images as it passes over the earth, acquiring data continuously and providing global coverage every 1-2 days. Fire detection is performed using an algorithm that exploits the strong emission of mid-infrared radiation from fires. However there is no information about the thermal anomaly that is detected - eg. it can be an agricultural fire, an urban fire or a flare from gas. A MODIS active fire detection represents the center of a 1km (approx.) pixel flagged as containing one or more actively burning hotspots/fires. It is not possible to determine the exact size of a fire represented by 1 pixel, but studies have shown that in good acquisition conditions a fire with a size of 1000$m^2$ can be reasonably detected (Giglio 2003).

### 2.3 Lights at night data

We have downloaded satellite imagery from the Visible Infrared Imager Radiometer Suite (VIIRS) sensor on board the NASA-NOAA Suomi NPP satellite (npp.gsfc.nasa.gov/). The data was acquired in April and October 2012 and is a global composite of cloud-free images utilizing the day-night band of VIIRS. For this research, we have used a subsampled image covering Cote d'Ivoire with a resolution of 3km per pixel (see Fig1.b).

## 3. Mobile phone data analysis in fire spots

### 3.1 Characterizing fires locations with night lights information and CDRs

We crossed the geographic location information from the available antennas and the fires. During the studied time interval, we identified 95 antennas at less than 1km distance of a fire, 81 antennas had a unique fire and 14 antennas were affected by two fires - making a total of 109 fires (see Fig1.c).

Based on the hypothesis that luminosity at night is correlated with population and economic activity, we set up a classification system devoted to automatically identifying the urban and rural areas using the lights at night information. With this system we classified the nature of the 95 places with both a mobile phone antenna and a fire during the studied period. For each one of the antenna positions, we integrated the intensity signal of the light at night imagery in an area with a 7.5 km radius. These values were clustered into 3 classes using a k-means algorithm. The resulting classes were composed by 8, 15 and 72 locations. A visual inspection of the elements of each class, both in the lights at night imagery and in Google Maps, revealed that the cluster with 8 locations corresponds to big cities (eg. Abidjan), the cluster with 15 elements signals small cities and the larger class corresponds to remote rural areas and some roads (see Fig1.f). In order to asses this type of classification – from urban to rural, we selected all the individual trajectories (from dataset *b*) that passed by a fire the same day of the fire detection. We found that for 18% of the antennas, no trajectory was logged that day. We do not know if it is because of the impact of the fire –ie. the network went down -, or if it is due to the low population sampling in the mobile phone data. From the remaining antennas with at least one person logging the day of the fire, we measured the number of individual trajectories for each cluster. The result shows a direct relationship between the number of trajectories and the urban-rural classification obtained from the lights data (see Fig.1d). This suggests that both mobile traffic and lights at night are proxies for the urban-rural classification of the fire locations.

In a second experiment, we explored the potential impact of fires at the antenna level (starting and terminating calls) at a short time scale of a few days. We added the number of calls per hour of all the antennas in each of the clusters after a temporal alignment with the fire date (t=0 is 12h of the day of the fire). The mobile phone activity is given by hour, while for the fire temporal resolution it is daily. When adding the activity of all the antennas of a category, in order to avoid over representation of a single antenna with many calls, we normalized the number of calls per hour of each antenna by the maximum number of calls in the same antenna during five days centered in the fire date. That is, we obtained the mean number of calls per hour from two days before a fire until two days after that fire (see Fig1.g). Interestingly, we observed that the day after the fire, there are more calls in the morning and less in the evening. In rural areas, the typical pattern of daily calls has a couple of peaks, one in the morning and one in the evening – being the latter of slightly higher intensity. We noticed that after a day with a fire event, the intensity of the two peaks was inverted, that is, more calls in the morning than in the evening. This might be because people are seeking information on what happened. Two days after the fire, the typical pattern is recovered, suggesting a rapid resumption of normal activity (at least in most of the cases). In the small city category, we observed a large increase of terminating calls in the morning of the day after the fire, which again might be associated to the requests for information attention-worthy events. In the big cites, our data indicate that fires result in a reduced number of calls.

We also tried to explore potential longer term effects that might be caused by devastating fires. However, we obtained very unstable results when examining patterns at a weekly-monthly time scale. We suspect that it is due to non-uniform calling patterns in the analyzed antennas, which brings an additional level of complexity. When examining the data at daily scale, we found three patterns that make any kind of temporal alignment around fires a difficult task (see Fig1.e). First there is an inherent pattern in the mobile phone activity of Orange customers in Cote d'Ivoire: there are many calls at the beginning of the month, with a decreasing drift until the end of the month. We think that this is pre-paid phone typical pattern – people have money at the beginning of the month so they top-up their sim cards. It would be interesting to know the proportion of pre-paid and contract subscribers for a deeper study of this effect. Second, during the period analyzed - from December to April 2012- we observed that for antennas in several regions there are fewer calls around Christmas. Third, there are some punctual spikes of activity, as the 1$^{st}$ of January or the 1$^{st}$ of March. All these factors point to an open question on how to aggregate the information from different antennas occurring at different times points -in our case when fire is detected – at a temporal scale in the order of weeks or months.

**Figure 1**

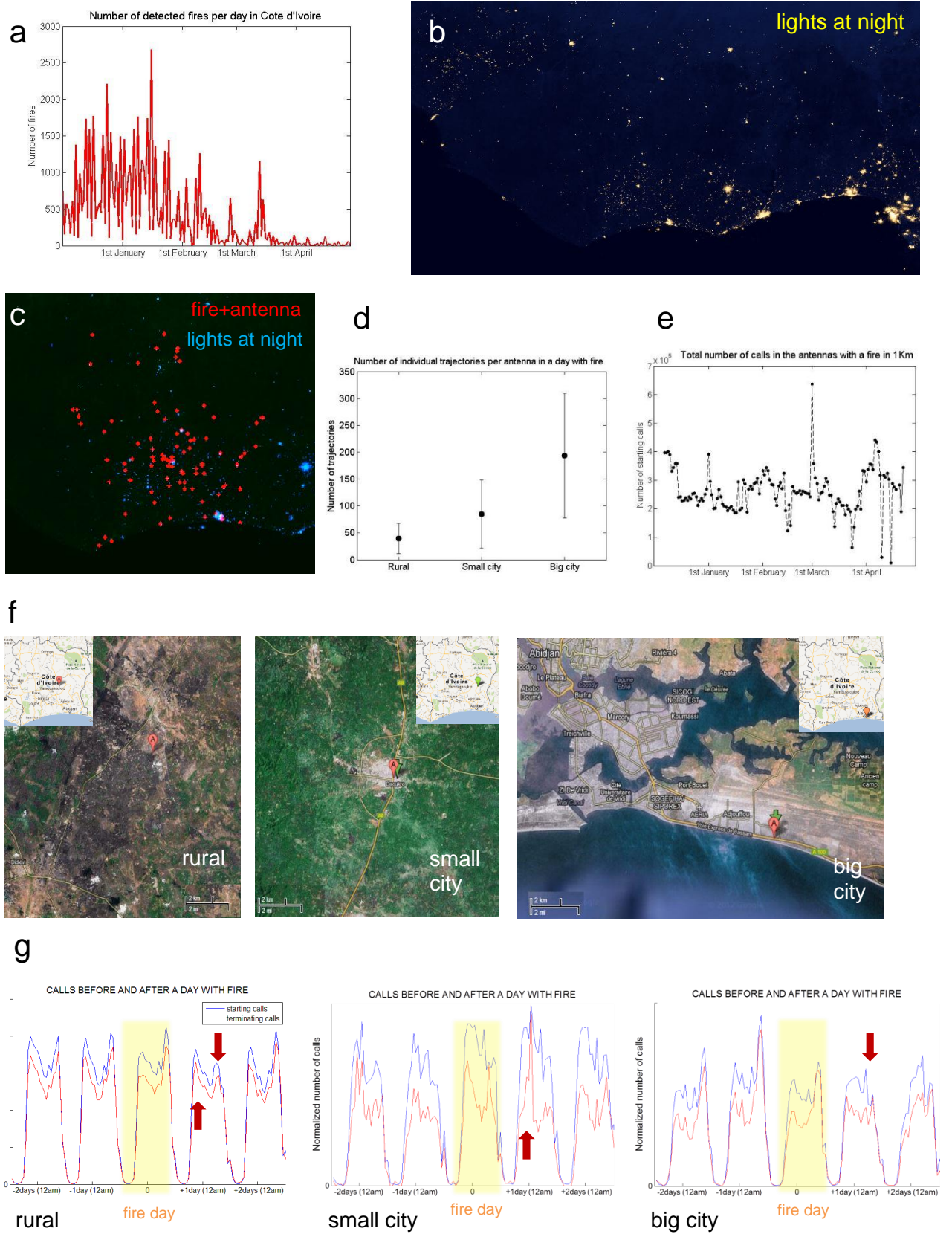

Fires, Lights and Mobiles Phones for Development?**3.2 Visualizing Dynamic Call Detail Records**

The previous analyses reinforced the idea that handling complex geo-localized information as mobile phone activity and fire data requires the development of dedicated visual analytics tools capable of showing the data in a comprehensive manner. There exist available ad-hoc visualization platforms that allow seeing general mobile traffic statistics at high scale (eg. www.geofast.net) and some non-interactive demonstrative videos (eg. www.villevivante.ch/). However, there is not a standard interactive visual analytics platform that allows to visually understanding individual trajectories of this nature with high spatio-temporal resolution. Because of this, we have developed the basic architecture of an interactive visualization platform focused on providing a multiscale and dynamic view of the individual trajectories and their spatial relations together with heterogeneous databases such as the detected fire locations. This platform - so-called MOBILOMICS (www.die.upm.es/im/archives/mobilomics/)- has been developed using *Processing* language (processing.org/) and the *Unfolding Maps* library (unfoldingmaps.org/) and allows us to load several maps from OpenStreetMap (licensed under the Open Data Commons Open Database License), mobile phone trackings and specific spatio-temporal events. With this platform, we can browse and explore the data in space and time at different scales, searching and selecting specific subsets of individual trajectories that share similar characteristics. As a visualization example, in Fig2, we see a capture that shows all the trajectories of people that logged the same day that a fire occurred in three different antennas corresponding to each of the 3 categories: rural – small city – big city.

**Figure 2**

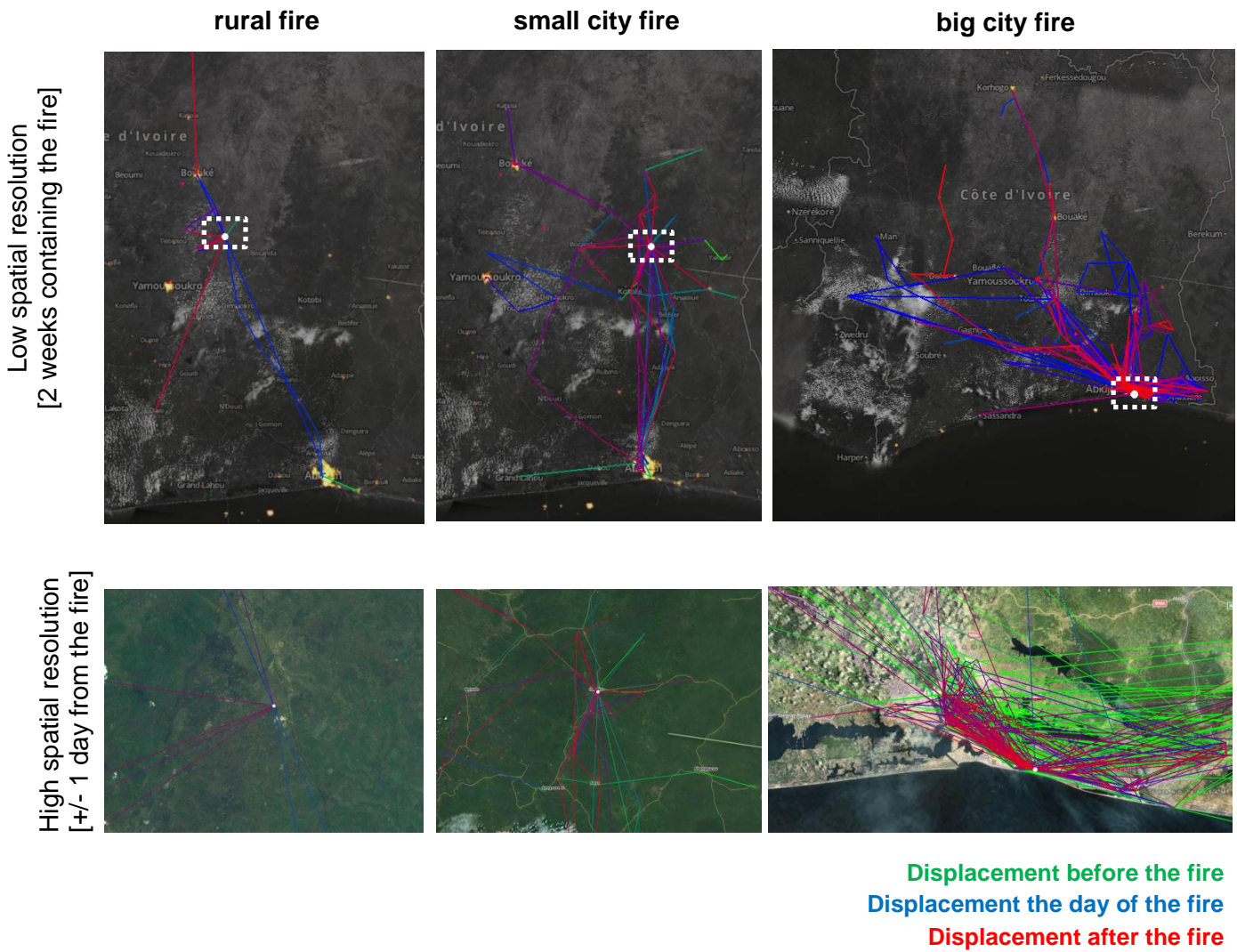

**Displacement before the fire**
**Displacement the day of the fire**
**Displacement after the fire**

## 4. Discussion and open questions

In this report we have presented the idea of merging fire, night lights and mobile phone information to explore behavioral fingerprints that might be useful for development. We have introduced some analytical methodologies, shown preliminary data analysis and developed an interactive visualization platform that allows integrating and exploring individual trajectories and fire data with high spatiotemporal resolution. While this research is in progress, and the preliminary discoveries neither are concluding nor statistically significant, we have better understood the kind of questions related to development that this data might answer, and we have found some useful clues encouraging further investigation. The relation between phone calls and light at night data suggests that the previous works relating light data with macroeconomic indicators as GDP might be improved by adding mobile data analytics to the mix, so that a finer scale both in time and scale becomes feasible to not only measure economic activity, but also to characterize urban and rural areas. Regarding the potential effect of fires in mobile phone activity – and potential proxy indicators of human behavior - , we have identified some changes in the patterns of calls at hourly scale just after a fire occurred. We hypothesize that this information – and the time needed to recover a normal calling pattern - might be used to track the recovery of a zone after a fire emergency. Further research should be done to understand the particular cases depending on the nature of the fire (eg. urban fire, agricultural related fire or conflict fire) that might be approached by crossing this output with emergency records or field surveys, and with potential use to optimize emergency resources and protocols. When analyzing the effect of fires at longer time scales – i.e weeks, we have identified some data biases that imply normalization challenges that should be investigated. However, these long term effects could be of high interest. For instance, in an urban environment, we might imagine some people forced to change their home or job location because of a fire. In a rural environment, we might imagine this impact translated into a decreased agricultural activity. At a broader scale we could infer possible economic deceleration due to devastating fires thanks to phone activity records -other official resources should be used to verify and extract robust characterization and classification of these social dynamics. For further research, it would be extremely useful to dispose of a longer dataset of mobile phone activity, as we suspect that there will be many antennas close to the almost 60000 fires identified during the 5 months studied. The effects of the subsampling of the mobile dataset remain unknown and it would be interesting to asses which subsample is representative and optimal for this kind of research. In future developments, we expect to include the data analysis methods and different ground truth data from the off-line world into the visualization interface, with special emphasis in the dynamic properties of the individual trajectories data in order to better understand the particular flows in and out the fire locations depending of the nature and consequences of the fire.

## Acknowledgments

The authors would like to thank Eva Kaplan and Constanza Blanco for their fruitful discussions and feedback on the study. This research was partially funded by the Picata program from the Moncloa Campus of International Excellence, Universidad Politécnica de Madrid and Universidad Complutense de Madrid, Spain.